\documentclass[conference]{IEEEtran}
\IEEEoverridecommandlockouts
\usepackage{cite}
\usepackage{amsmath,amssymb,amsfonts}
\usepackage{algorithmic}
\usepackage{graphicx}
\usepackage{textcomp}
\usepackage{multirow}
\usepackage{threeparttable}
\usepackage{diagbox}
\usepackage{xcolor}
\def\BibTeX{{\rm B\kern-.05em{\sc i\kern-.025em b}\kern-.08em
    T\kern-.1667em\lower.7ex\hbox{E}\kern-.125emX}}
\begin{document}

\title{Design and Evaluation Frameworks for Advanced RISC-based Ternary Processor\\
\thanks{Identify applicable funding agency here. If none, delete this.}
}

\author{\IEEEauthorblockN{Dongyun Kam, Jung Gyu Min, Jongho Yoon, Sunmean Kim, Seokhyeong Kang and Youngjoo Lee}
\IEEEauthorblockA{Department of Electrical Engineering \\
Pohang University of Science and Technology, Pohang, South Korea \\
\{rkaehddbs,mjg1104,yoonjongho99,sunmean,shkang,youngjoo.lee\}@postech.ac.kr
\\
}
}

%double-blind 
%\author{\\ \\
%\\ \\}
%\maketitle

\maketitle

\begin{abstract}
In this paper, we introduce the design and verification frameworks for developing a fully-functional emerging ternary processor.
%ART-9 core, a fully-functional 9-trit advanced RISC-based ternary processor.
Based on the existing compiling environments for binary processors, for the given ternary instructions, the software-level framework provides an efficient way to convert the given programs to the ternary assembly codes.
%using 24 custom ternary instructions, resulting in a comparable program size to the baseline 32-bit binary processor.
%Several simulation environments including the ternary gate-level simulator and the cycle-accurate instruction-set simulator are additionally developed to test the proposed ART-9 core with benchmark programs.
We also present a hardware-level framework to rapidly evaluate the performance of a ternary processor implemented in arbitrary design technology.
As a case study, the fully-functional 9-trit advanced RISC-based ternary (ART-9) core is newly developed by using the proposed frameworks. 
Utilizing 24 custom ternary instructions, the 5-stage ART-9 prototype architecture is successfully verified by a number of test programs including dhrystone benchmark in a ternary domain, achieving the processing efficiency of 57.8 DMIPS/W and $\textbf{3.06} \times \textbf{10}^{\textbf{6}}$ DMIPS/W in the FPGA-level ternary-logic emulations and the emerging CNTFET ternary gates, respectively.
\end{abstract}

\begin{IEEEkeywords}
Ternary processor, Instruction set architecture, RISC, Emerging computer design, Multi-valued logic circuits
\end{IEEEkeywords}

\section{Introduction}
The dimensional down-scaling of CMOS technology has been continuously focused on increasing hardware efficiency of digital circuits \cite{yeap2013smart}.
However, the performance improvement from the recent down-scaling is now expected to meet the potential limitation especially caused by the increased interconnecting/routing overheads \cite{magen2004interconnect}. 
Among the different solutions to address this limitation, the multi-valued logic (MVL) circuits have been recently gaining great popularity due to their attractive potential for reducing the circuit-level complexity as well as the routing burden even at the aggressive down-scaling technology \cite{gaudet2016survey}.
For the practical implementation of the ternary-based system, which is a starting point of MVL solutions, numerous technologies have been proposed to solve the stability issue of each voltage/current level with emerging devices including carbon nanotube FETs (CNTFETs) \cite{lin2009novel}, graphene barristors \cite{yang2012graphene}, and CMOS-based ternary transistors \cite{shin2015compact}.
In general, the prior studies on ternary circuits mainly present the potential expectations of gate-level performances \cite{kim2018optimal,kim2020logic}.
For circuit-level studies, some digital building blocks such as adder, multiplier, and flip-flop have been also investigated to extend the concept of gate-level evaluation in ternary domains \cite{heo2018ternary,kang2017novel,choi2021design}.
Due to the lack of systematic design-level strategies, on the other hand, the system- or processor-level explorations for ternary-based digital solutions are rarely reported in the open literature with few details \cite{PyconTernary}.

In this work, we introduce advanced design and evaluation frameworks to realizing ternary processors, measuring actual performances with the practical benchmark programs. %for the first time.
In contrast that the previous studies only present limited concepts to only test processing blocks in ternary number systems \cite{DouglasW,narkhede2013design}, we develop a 9-trit advanced RISC-based ternary (ART-9) core by adopting the proposed frameworks, presenting the fully-functional top-level ternary microprocessor.
%designed by using the in this work defines the top-level architectures for realizing a fully-functional RISC-based ternary microprocessor.
Based on the 9-trit instruction-set architecture (ISA) with 24 custom ternary instructions, more precisely, the proposed software-level framework provides an efficient way to convert the existing binary programs to the ternary codes, even reducing the program size compared to the baseline codes with RV-32I ISA \cite{RISC-V}.
The hardware-level framework offers the cycle-accurate simulator and the technology mapper, providing the quantitative evaluations of the pipelined ART-9 architecture for arbitrary design technology. 
Targeting the specialized 5-stage pipelined architecture, as a case study, the proposed ART-9 core achieves the processing efficiency of 57.8 DMIPs/W and $3.06 \times 10^6$ DMIPs/W when we use the FPGA-level ternary-logic emulations and the emerging CNTFET-based ternary gates \cite{kim2020logic,kim2018optimal}, respectively, reporting the first full-level evaluations of ternary processors.

\section{Background}
\subsection{Ternary Number Systems}
Based on the integrated ternary gates using three voltage levels such as GND, VDD/2 and VDD, the ternary circuits are dedicated to the arithmetic operations in a ternary number system.
Like the binary case, there are in general two types of ternary fixed-point number systems: unsigned and signed systems.
For an $n$-trit number $X = (x_{n-1},x_{n-2},...,x_0)_3$, where $x_i \in \{ 0, 1, 2 \}$, the unsigned ternary number only represents positive integers from 0 to $3^{n}-1$ by interpreting an $n$-trit sequence into a decimal value $Y_{\text{unsigned}}$ as follows. 
\begin{equation}
    Y_{\text{unsigned}} = \sum^{n-1}_{k=0} x_k3^k.
\end{equation}

Although the unsigned number system is useful for denoting indices of general-purposed ternary registers (GPTR) and addresses of ternary instruction/data memories (TIM/TDM), it is obviously required to support the signed arithmetic operations for performing the general data processing.
Therefore, the singed ternary number system is considered to interpret the given $n$-trit sequence into the negative value.
Among different ways to develop the ternary signed numbers \cite{DouglasW}, in this work, we adopt the balanced signed number system, where each trit is now an element from the balanced set, i.e, $x_i \in \{ -1, 0, 1 \}$ \cite{DouglasW}.
Then, a numerical value $Y_{\text{signed}}$ of $n$-trit ternary number $X$ is still calculated in the same way as the unsigned representation shown in (1).
%\begin{equation}
%    Y_{\text{signed}} = \sum^{n-1}_{k=0} (x_k-1)3^k,
%\end{equation}
Compared to the unbalanced approaches in \cite{DouglasW}, it is reported that the arithmetic operations in balanced ternary numbers can be simplified according to the conversion-based negation property \cite{kim2020logic,narkhede2013design}.
To develop the proposed frameworks for general-purposed ternary processors, therefore, the balanced representation is definitely suitable by requiring fewer ternary gates for the practical realization \cite{kim2020logic}.

\begin{figure} [!t]
    \centering
    \includegraphics[scale=1.0]{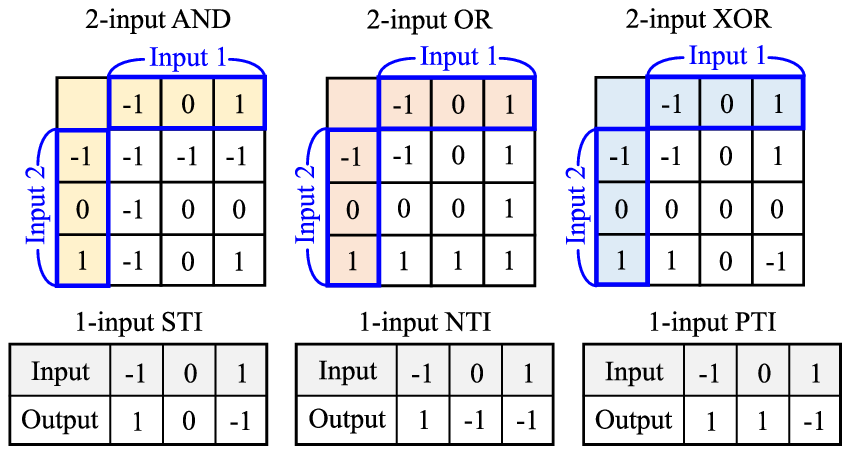}
    \caption{Truth tables of ternary logic operations.}
    \label{fig.fig1}
%    \vspace{-3mm}
\end{figure}

\subsection{Ternary-based Arithmetic and Logical Operations}
Similar to basic operations of RISC-based binary processors \cite{RISC-V}, the ternary processor should support logic and arithmetic operations to perform the general user-level programs.
As reported in \cite{DouglasW}, the balanced ternary logic operations include AND, OR, XOR, standard ternary inverting (STI), negative ternary inverting (NTI) and positive ternary inverting (PTI), where the detailed truth tables are exemplified in Fig. 1.
%Fig. 1 shows truth tables for the detailed computations of the ternary logic operations.
Compared to the familiar two-input logic gates such as AND, OR, and XOR, note that the inverting operation consists of three fundamental functions (denoted as STI, NTI, PTI in Fig. 1), considering as the most important processing in the balanced ternary number system \cite{DouglasW}.
It is also possible to define the proper two-input arithmetic operations, which are comparable to the well-established functions in the binary processor \cite{RISC-V}.
For example, the fundamental functions including ternary addition, comparison, multiplication, and division, have been extensively studied for the next-generation computer arithmetic \cite{parhami2013arithmetic,heo2018ternary,kang2017novel }.
%For handling the hardware-level issues, it has been revealed that the ternary addition and comparison can be implemented by adopting the tree-like architecture \cite{kim2020logic, parhami2013arithmetic}.
Utilizing the negation operations in Fig. 1, it is also possible to simply utilize the ternary subtraction based on the pre-designed ternary adder \cite{parhami2013arithmetic}.
%The recent researches have also presented design-level approaches for the ternary multiplication and division, which may improve the multimedia performance at the advanced processor architectures \cite{kim2021low,parhami2013arithmetic}. 

\begin{figure} [!t]
    \centering
    \includegraphics[scale=1.0]{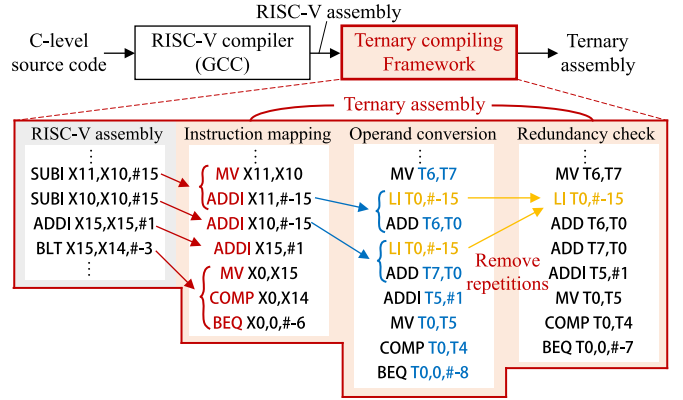}
    \caption{The proposed software-level compiling framework.}
    \label{fig.fig2}
\end{figure}

\section{Proposed Frameworks for Ternary Processors}

\subsection{Software-Level Compiling Framework}

For the full-level ternary processor implementation, based on a given ternary ISA, it is important to prepare ternary-based assembly (or executable) programs in an easy way.
With the assistance of the existing RISC-V tool chains in open-source domains \cite{8714897}, in this work, we first develop a software-level compiling framework supporting instruction mapping, operand conversion, and redundancy checking, which can efficiently generate the ternary assembly benchmarks for arbitrary C-based source codes. 
Fig. 2 conceptually illustrates the processing steps of the proposed software-level framework.
Note that the input C-level program is firstly handled by an open-source compiler for RV-32I ISA, obtaining an assembly sequence of 32-bit instructions. 
Then, the instruction mapping step is activated to translate the 32-bit instructions into pre-defined ternary instructions.
For a binary instruction that cannot be directly converted with a ternary version, we utilize several primitive sequences of ternary instructions, still offering valid mapping results by allowing a few more instructions.
After mapping ternary instructions, the operand conversion step is followed to find the ternary representations of immediate values in the baseline binary instructions. 
Depending on the definition of ternary instructions, it might be required to add more instructions to construct the large-sized operands in ternary number systems.
Note that the operand conversion step also supports the register renaming when the given ternary ISA uses fewer general-purposed registers than the baseline binary processor. 
As the mapping and conversion steps may utilize additional instructions, the final redundancy checking phase finds the meaningless instructions by investigating the duplicated operations, removing them to minimize the overall code size. 
During the elimination of redundant instructions, the proposed framework also re-calculates the branch target addresses to ensure the correct results. 
As the proposed software-level framework is based on the well-established compiling environments for binary processors, we can purely focus on increasing the mapping quality in the ternary domain by deeply considering the characteristics of ternary instructions, relaxing the development efforts of ISA-dependent processor-design frameworks.
Targeting the proposed ART-9 ISA, as a case study, the proposed software-level framework easily generates various ternary codes with reduced memory requirements, even saving the program size of dhrystone benchmark by 54\% compared to the baseline processor of RV-32I ISA \cite{RISC-V}.
%With the RISC-V assembly generated from the RISC-V compiler, the binary-to-ternary converter newly constructs the atched ART-9 assembly codes by considering the number of general-purposed registers and the range of immediate values.
%Note that if we cannot completely replace a RISC-V instruction into a ART-9 instruction, it is allowed to utilize ultiple ART-9 instructions.
%For instance, as shown in Fig.2, the red-highlighted RISC-V instruction is converted into three ART-9 instructions due o supporting few number of registers and operands.
%To minimize the converting program size, we optimize the number of instructions by removing the existing repeated odes, which unnecessarily perform the same operations.
%To evaluate the proposed ART-9 processor in a cycle-accurate instruction set simulator or a gate-level simulator, the ART-9 assembly simply translates the optimized codes into the machine-level codes that are encoded by trit values.

\begin{figure} [!t]
    \centering
    \includegraphics[scale=1.0]{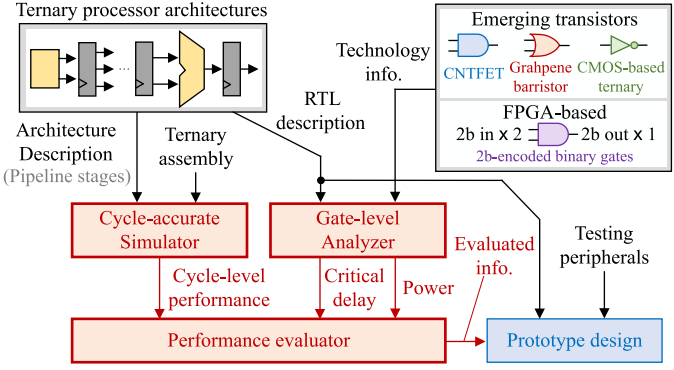}
    \caption{The proposed hardware-level evaluation frameworks.}
    \label{fig.fig3}
\end{figure}

\subsection{Hardware-Level Evaluation Framework}

Using ternary assembly codes, which can be obtained by the proposed software-level compiling framework, we develop the hardware-level evaluation framework allowing an efficient way to develop the prototype of ternary processor. 
As illustrated in Fig. 3, the hardware-level framework includes a cycle-accurate simulator, a gate-level analyzer, and a performance estimator.
The cycle-accurate simulator accepts the high-level description of the pipelined ternary processor, and provides the required processing cycles for performing the input ternary assembly codes.
With the synthesizable RTL design corresponding to the high-level architecture description, the proposed gate-level analyzer can estimate the critical delay as well as the power consumption of ternary processor. 
Note that we define the property description of the design technology as another input of gate-level analyzer, which includes delay and power characteristics of primitive building blocks, enabling more accurate analysis results depending on the target technology.
As depicted in Fig. 3, the performance estimator gathers all the outputs from prior steps, and finally generates the overall evaluation information of the ternary processor implemented in certain design technology. 
%If we can accept the hardware-level evaluation results from the proposed framework, then it is now acceptable for actually implement the prototype ternary processor with pre-designed pheri.
By utilizing multiple evaluation steps and even considering the technology-oriented information, before starting the actual implementation phase with pre-designed peripherals, we can remarkably reduce the design efforts of the custom ternary processor with the proposed hardware-level evaluation framework.

%}For the processing verification of the fully converted machine level codes, we present a cycle-accurate instruction simulator that considers an arbitrary processor architecture supporting the proposed ART-9 instruction set.
%The proposed instruction-level simulator shows changes of processed values in the assumed r}egisters and memory for every instructions ,and also provides the required number of cycles f}or completing each program by considering the inserted stalls at the given pipelined a}rchitecture.
%}
%\textcolor{blue}{
%}For the logic-level verification and performance evaluation, in addition, we emulated the primitive ternary gates in Verilog-HDL language, mapping the balanced trit values $(-1, 0, +1)$ with the existing symbols $(0, z, 1)$.
%}It is possible to design all the internal modules with the primitive ternary gates, which include pre-desiend building blocks such as ternary registers, ternary multiplexers, TRF, T}IM, and TDM \cite{kim2020logic,choi2021design,mohammaden2021cntfet,kim2020extreme}.
%The proposed technology mapper applies information of the given technology library to V}erilog-described architecture, finally providing power consumption and operating frequency.
%Note that it is possible to support the commercialized binary FPGA-based technology as well as CNTFET-based ternary technology by encoding each trit value into a 2-bit value.
%}

\begin{figure*}[ht]
    \centering
    \includegraphics[scale = 1.0]{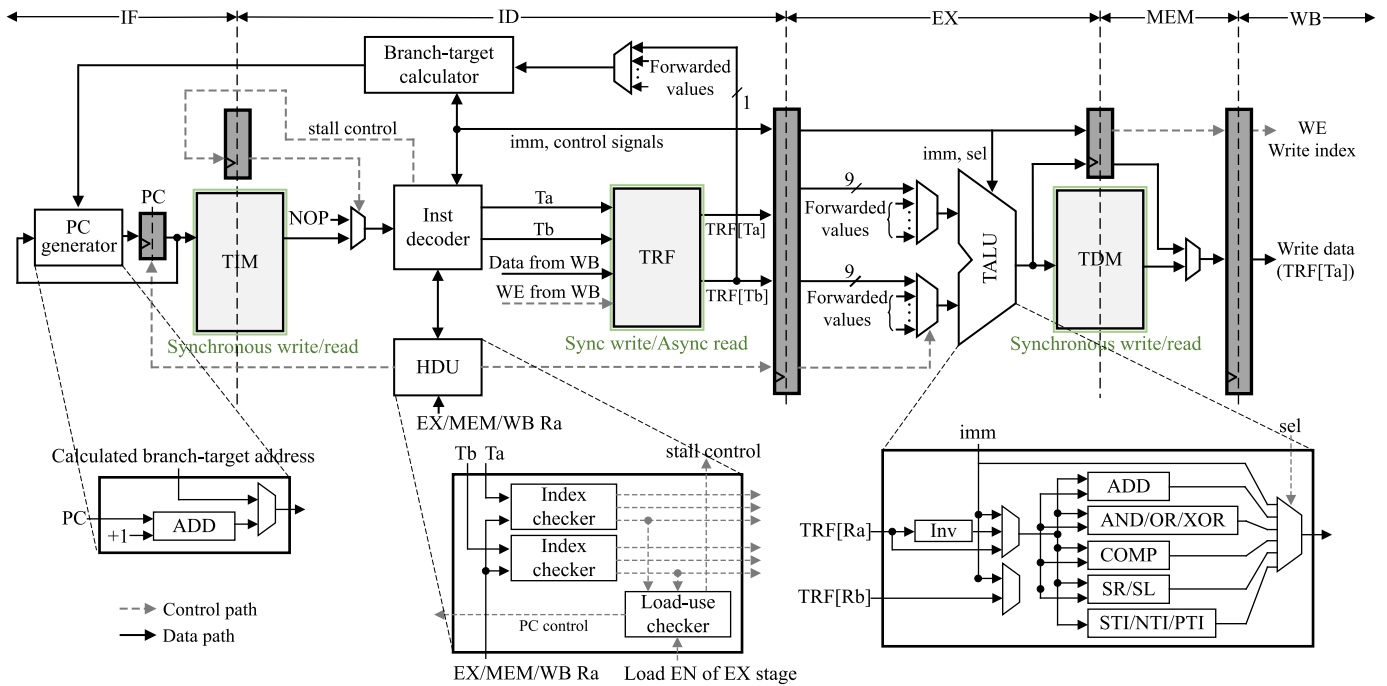}
    \caption{5-stage pipelined architecture of the proposed ART-9 processor.}
    \label{fig:fig4}
\end{figure*}

\begin{table}[!t]
\centering
\renewcommand{\arraystretch}{1.3}
\renewcommand{\tabcolsep}{1.1mm}
\caption{Summary of ART-9 Ternary Instructions}
\begin{tabular}{|cc|c|c|}
\hline
\hline
%8 & 7 & 6 & 5 & 4 & 3 & 2 & 1 & 0 & \\ \hline
%\multicolumn{2}{|c|}{\diagbox[width = 10em, height = 2.9em]{Instruction}{Description}} & \textbf{8} & \textbf{7} & \textbf{6} & \textbf{5} & \textbf{4} & \textbf{3} &\textbf{2} & \textbf{1} & \textbf{0} & \textbf{Operation} & \textbf{Assembly usage} \\ \hline 
\multicolumn{1}{|c|}{Type} & 9-trit instructions & Operation \\ \hline
%\textbf{Type}&\textbf{Mnemonics} & \multicolumn{3}{|c|}{\textbf{Func3/imm}} & \multicolumn{2}{c|}{\textbf{Tb/imm}} & \multicolumn{2}{c|}{\textbf{Ta}} & \multicolumn{2}{c|}{\textbf{Opcode}} &\textbf{Operation} & \textbf{Assembly usage}\\ \hline 
\multirow{12}{*}{R} & \multicolumn{1}{|c|}{MV Ta,Tb} & TRF[Ta] = TRF[Tb] \\ \cline{2-3}
& \multicolumn{1}{|c|}{PTI Ta,Tb}& TRF[Ta] = PTI(TRF[Tb])  \\ \cline{2-3}
& \multicolumn{1}{|c|}{NTI Ta,Tb}&  TRF[Ta] = NTI(TRF[Tb])  \\ \cline{2-3}
& \multicolumn{1}{|c|}{STI Ta,Tb}& TRF[Ta] = STI(TRF[Tb]) \\ \cline{2-3}
& \multicolumn{1}{|c|}{AND Ta,Tb}&  TRF[Ta] = TRF[Ta] $\&$ TRF[Tb]  \\ \cline{2-3}
& \multicolumn{1}{|c|}{OR Ta,Tb}&  TRF[Ta] = TRF[Ta] $|$ TRF[Tb]\\ \cline{2-3}
& \multicolumn{1}{|c|}{XOR Ta,Tb} & TRF[Ta] = TRF[Ta] $\oplus$ TRF[Tb]\\ \cline{2-3}
& \multicolumn{1}{|c|}{ADD Ta,Tb}& TRF[Ta] = TRF[Ta] $+$ TRF[Tb]  \\ \cline{2-3}
& \multicolumn{1}{|c|}{SUB Ta,Tb}&  TRF[Ta] = TRF[Ta] $-$ TRF[Tb] \\\cline{2-3}
& \multicolumn{1}{|c|}{SR Ta,Tb}&  TRF[Ta] = TRF[Ta] $\gg$ TRF[Tb][1:0]   \\ \cline{2-3}
& \multicolumn{1}{|c|}{SL Ta,Tb}& TRF[Ta] = TRF[Ta] $\ll$ TRF[Tb][1:0]  \\ \cline{2-3}
& \multicolumn{1}{|c|}{COMP Ta,Tb}&  TRF[Ta] = $compare$(TRF[Ta],TRF[Tb])  \\ \hline \hline
\multirow{6}{*}{I} & \multicolumn{1}{|c|}{ANDI Ta,imm} & TRF[Ta] = TRF[Ta] $\&$ imm[2:0] \\ \cline{2-3}
& \multicolumn{1}{|c|}{ADDI Ta,imm}&  TRF[Ta] = TRF[Ta] $+$ imm[2:0]  \\ \cline{2-3}
& \multicolumn{1}{|c|}{SRI Ta,imm}&  TRF[Ta] = TRF[Ta] $\gg$ imm[1:0]  \\ \cline{2-3}
& \multicolumn{1}{|c|}{SLI Ta,imm}& TRF[Ta] = TRF[Ta] $\ll$ imm[1:0]  \\ \cline{2-3}
& \multicolumn{1}{|c|}{LUI Ta,imm}&  TRF[Ta] = \{imm[3:0],00000\}  \\ \cline{2-3} 
& \multicolumn{1}{|c|}{LI Ta,imm}&  TRF[Ta] = \{TRF[Ta][8:5],imm[4:0]\}  \\ \hline \hline
\multirow{4}{*}{B} & \multicolumn{1}{|c|}{BEQ Ta,B,imm} &  PC = PC $+$ imm[3:0] if TRF[Tb][0] == B \\ \cline{2-3}
& \multicolumn{1}{|c|}{BNE Ta,B,imm}&  PC = PC $+$ imm[3:0] if TRF[Tb][0] != B  \\ \cline{2-3}
& \multicolumn{1}{|c|}{JAL Ta,imm}&  TRF[Ta] = PC$+$1, PC = PC $+$ imm[4:0] \\ \cline{2-3}
& \multicolumn{1}{|c|}{JALR Ta,Tb,imm}&  TRF[Ta] = PC$+$1, PC = TRF[Tb]$+$imm[2:0] \\ \hline \hline
\multirow{2}{*}{M}& \multicolumn{1}{|c|}{LOAD Ta,Tb,imm}&  TRF[Ta] = TDM[TRF[Tb]+imm[2:0]  \\ \cline{2-3}
& \multicolumn{1}{|c|}{STORE Ta,Tb,imm}& TDM[TRF[Tb]+imm[2:0]] = TRF[Ta]  \\ \hline
\hline
\end{tabular}
\end{table}

\section{ART-9 Core Design for Proposed Frameworks}

\subsection{ART-9 Instruction Set Architecture}

Based on the balanced ternary number systems, the proposed ART-9 processor defines 9-trit-length ISA following the properties of contemporary RISC-type binary processors \cite{RISC-V}.
Table I summarizes 24 ART-9 ternary instructions processing 9-trit data values, which are the essential inputs at the proposed software-level compiling framework.
%, where the undefined encoding patterns are reserved for the further extension. 
By matching the word length of both instruction and data, we can allow the regular structure for realizing TIM and TDM.
To fetch an instruction by accessing the TIM, we use a special-purposed 9-trit register, i.e., the program counter (PC) register containing the instruction address.
In order to store the intermediate data, like the modern processor architectures \cite{RISC-V,yiu2015definitive}, the ART-9 core also includes a ternary registerfile (TRF) including nine general-purposed registers, each of which is accessed by using a 2-trit value.
Utilizing the load-store architecture used for typical RISC processors \cite{schuiki2020stream}, there are four instruction categories in ART-9 ISA; R-type, I-type, B-type, and M-type.

For the R-type instructions, considering the recent studies \cite{li2019reduce}, we select essential 12 logical/arithmetic functions as depicted in Table I.
In fact, most R-type instructions are typical two-address instructions, which fetch two 9-trit operands in TRF, whose 2-trit indices are denoted as Ta and Tb, and then overwrite a 9-trit result to the register TRF[Ta].
Note that some R-type instructions specialized for inversion and data-movement operations use only one source operand from Tb, where the destination operand is still Ta to have the regular encoding patterns, relaxing the complexity to decode the fetched instruction. 
In addition, we also realize an R-type comparison instruction (COMP), where the least significant trit (LST) of the destination register TRF[Ta] denotes the comparison result of two input operands with the dedicated function $compare()$ in Table I. 
More specifically, the LST of TRF[Ta] is set to be zero when the two inputs are the same, otherwise it becomes $+1$ (or $-1$) if TRF[Ta] $>$ TRF[Tb] (or TRF[Ta] $<$ TRF[Tb]).
This COMP instruction plays an important role to improve the code density by allowing the conditional execution of the following branch instructions.

In order to reduce the generation complexity of constant values, the technique to encode immediate values into the instruction is generally used for reducing the size of user-level programs \cite{RISC-V,yiu2015definitive}. 
The proposed ART-9 processor also supports immediate-based processing with I-type instructions. 
As described in Table I, unlike the R-type instructions offering various functions, we only allow immediate values at addition, AND logic, and shift functions, which are known to be the most common operations in practice \cite{li2019reduce}. 
Due to the limited trit-width for denoting an embedded immediate, there could be extra overheads to realizing full-length (9-trit) constant values. 
Instead of utilizing a series of the shift-and-addition process to store a 5-trit immediate value initialized by a \emph{load immediate} (LI) instruction, we adopt a special I-type instruction named \emph{load upper immediate} (LUI), which is introduced at the RISC-type processors for making a large-sized constant value \cite{RISC-V, yiu2015definitive}.
As a result, the ART-9 ISA offers an acceptable flexibility to use wide ranges of immediate values, suitable for the resource-limited processing environments.

Besides the logical/arithmetic instructions, it is also required to define the branch-related instructions changing the PC value, which is denoted as B-type instructions in Table I.
In the proposed ART-9 cores, we introduce four B-type instructions including two conditional branch operations associated with the PC-relative addressing, which are referred to as BEQ and BNE as shown in Table I.
To utilize these conditional operations, as described earlier, we preset the LST of TRF[Tb] in BEQ or BNE by using a COMP instruction, so that a 1-trit B value in BEQ or BNE is compared to check the branch-taken condition.
In addition, we define two unconditional jump-and-link instructions (JAL and JALR), which are mainly used for the subroutine calls.
Adopting the PC-relative addressing, similar to the conditional branches, the JAL instruction uses the PC value as a base address added by a 5-trit immediate.  
On the other hand, by using the JALR instruction, we can use the stored 9-trit value in TRF to set the base address with a small-sized 3-trit immediate, allowing more long-range jumps. 
As depicted in Table I, note that this base-register addressing is also used to access TDM with M-type load/store instructions, reducing the hardware complexity with the shared datapath.

%\section{\textcolor{blue}{Hardware-level Evaluation Frameworks}}

\subsection{5-stage Pipelined ART-9 Architecture}

To support the proposed ART-9 ISA efficiently, we develop in this work a simple but efficient pipelined architecture, which is used for input descriptions of the proposed hardware-level evaluation framework.
As shown in Fig. 4, similar to the lightweight RISC-type designs \cite{schuiki2020stream}, there are five stages for fetching the instruction from TIM (IF), decoding the fetched instruction (ID), executing the arithmetic/logical operations (EX), accessing the TDM (MEM), and updating the result to TRF (WB).
The ternary pipelined registers are newly developed to keep the results from each stage, making a balanced pipelined processing.
We also introduce the synchronous single-port TIM and TDM designs for reducing the memory-accessing latency, where the TRF in this work supports two asynchronous read ports and one synchronous write port.
The ternary arithmetic logic unit (TALU) in EX stage is specialized to perform various operations depending on the control signals from main decoder in ID stage.
In the pipelined ART-9 core, the hazard detection unit (HDU) in ID stage compares the adjacent instructions to determine the generation of hardware-level stall controls at the run time. 
For reducing the number of unwanted stalls as many as possible, we actively apply the forwarding multiplexers to get the correct 9-trit inputs at TALU, solving ALU-use data hazards.
To minimize the number of stalls from B-type instructions causing control hazards, the pipelined ART-9 processor utilizes the dedicated branch-target calculator as well as the condition checker in ID stage, directly forwarding the calculated address to update the PC register.
For checking the branch-taken conditions, in addition, forwarding one-trit values successfully mitigates the long and complex datapath starting from TRF, still allowing one-cycle stall after B-type instructions without increasing the overall critical delay.
As a result, we only observe the hardware-inserted stall cycles when there exist load-use data hazards and taken branches.
After detecting the stall insertion case, the main decoder at ID stage generates a stall control signal, which will be used for selecting the no-operation (NOP) at the next ID stage as shown in Fig. 4. 
Without introducing a dedicated NOP encoding, note that the proposed ART-9 ISA uses an ADDI instruction to denote the NOP operation with a zero-valued immediate.

\begin{figure} [!t]
    \centering
    \includegraphics[scale=1.0]{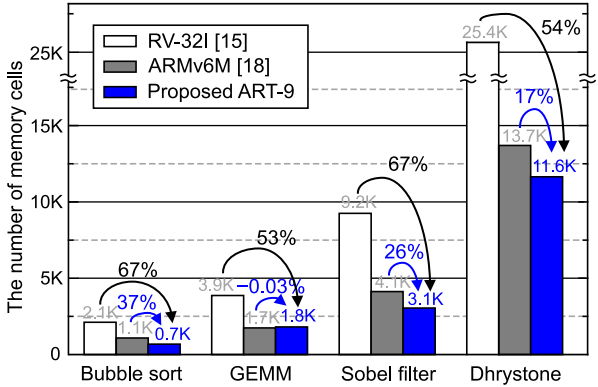}
    \caption{The number of memory cells for storing benchmark programs.}
    \label{fig.fig5}
\end{figure}

\section{Evaluation and Implementation Results}
\subsection{Benchmark Evaluations}

By using the proposed compiling framework, targeting the ART-9 ISA shown in Table I, we designed several ternary benchmark programs including the general computing algorithms; bubble sort, general matrix multiplication (GEMM), and sobel filter \cite{wang2012compression,zulehner2019matrix}.
In addition, for the first time, a dhrystone benchmark is also described with the ternary instructions by converting the existing dhrystone code of the RV-32I ISA, which is popularly used for evaluating the computing performance of general-purposed CPU cores \cite{york2002benchmarking}.
Fig. 5 depicts the effectiveness of the proposed ART-9 ISA by evaluating the required memory size for storing each benchmark program, showing that the proposed software-level framework offers acceptable assembly codes compared to the binary versions.
Note that we counted the number of memory cells for this comparison, as the ternary instructions necessitate the dedicated ternary memory where a storing cell can keep up to three different charge distributions \cite{choi2021design}.
Although we developed a simple 9-trit ISA including only 24 instructions, it is clear that the proposed ART-9 processor requires a much smaller memory size when compared to the binary counterparts; RV-32I using 32-bit instructions \cite{RISC-V} and ARMv6M with 16-bit instructions \cite{yiu2015definitive}.
If we consider implementation results of dhrystone codes, due to the short-length ternary instructions associated with the efficient software-level framework, for example, our ART-9 core reduces the number of required memory cells by 54\% and 17\% when compared to the RV-32I \cite{RISC-V} and ARMv6M \cite{yiu2015definitive} processors, respectively. 
%for example, our ART-9 core uses 1,294 instructions whereas the RV-32I processor utilizes 795 instructions.
%Although the proposed work necessitates more instructions to describe the dhrystone benchmark, as illustrated in Fig. 3, the total number of required memory cells can be even reduced by adopting the short-length instructions with the custom ternary encoding. 
In other words, there exist reasonable benefits of exploiting the ternary number systems, which can reduce the memory overheads while providing a similar amount of flexibility to binary ISAs with long-length instructions.

\begin{table}[!t]
\renewcommand{\arraystretch}{1.3}
\renewcommand{\tabcolsep}{1.1mm}
\centering
\caption{Simulation Results of Dhrystone Benchmark}
\label{table:functions}
\begin{tabular}{c|c|c|c}
\hline
\hline
& This work & VexRiscv \cite{VexRiscv} & PicoRV32 \cite{PicoRV32} \\ \hline
ISA Architecture & ART-9 ISA & RV-32I & RV-32IM \\ \hline
\# of instructions  & 24 & 40 &  48\\ \hline
Pipelined stages & 5 & 5 & 1\\ \hline
Multiplier & X & O & O \\ \hline
%\# of assembly & 1294 & 742 &  795\\ \hline
DMIPS/MHz & 0.42 & 0.65 &  0.31\\ \hline
\# of memory-cells & 11.6K trits & 25.4K bits & 23.7K bits\\ \hline
\hline
\end{tabular}

\end{table}

\begin{table}[!t]
\renewcommand{\arraystretch}{1.3}
\renewcommand{\tabcolsep}{1.1mm}
\centering
\caption{Processing Cycles for Different Test Programs}
\label{table:functions3}
\begin{tabular}{c|c|c|c|c}
\hline
\hline
 & Bubble sort & GEMM & Sobel filter & Dhrystone \\ \hline
This work & 2,432 & 10,748 & 7,822 & 134,200\\ \hline
PicoRV32 \cite{PicoRV32} & 9,227 & 11,290 & 18,250 & 186,607\\ \hline
\hline
\end{tabular}
\end{table}

\subsection{Hardware-level Evaluation Results}

Table II precisely shows evaluation results of the cycle-accurate simulator of different RISC-based processors running the dhrystone benchmark.
Note that our ART-9 core achieves 0.42 DMIPS/MHz by utilizing only 24 instructions, whereas the fully-optimized VexRiscv \cite{VexRiscv} and PicoRV32 \cite{PicoRV32} processor provides 0.65 and 0.31 DMIPS/MHz with more instructions and the dedicated multiplier, respectively.
Utilizing the optimized codes from the proposed compiling frameworks, the prototype ART-9 core can be designed with the comparable processing speed and much smaller size of memory cells.
In addition, Table III also shows that the proposed compiling framework efficiently optimizes the number of instructions for the other benchmarks.
As it is considered that the memory in general dominates the overall system complexity, the prototype ART-9 core offers the low-complexity computing platform even compared to the recent lightweight PicoRV32 processor with non-pipelined architecture. 
%which targets simple computing platform with one-stage pipelined architecture.
In other words, the optimized ART-9 ISA and processor, which utilize the proposed frameworks, successfully offer a reasonable solution achieving both the fast computing and the low hardware-cost, presenting the fully-functional processor design in the ternary domain.
%\subsection{\textcolor{blue}{Implementation Results}}

Using the proposed gate-level analyzer, we finally present implementation results of ART-9 prototypes.
For the simplified 32nm CNTFET ternary models without considering the parasitic capacitance \cite{kim2020logic}, we first estimated the gate-level costs of the 5-stage pipelined core as shown in Table IV.
The datapath of ART-9 core only required 652 standard ternary gates, consuming 42.7 $\mu$W when the operating voltage is set to 0.9 V.
According to the dhrystone result shown in Table II, the CNTFET-based ART-9 core achieves $3.06 \times 10^6$ DMIPS/W, showing that the emerging ternary device leads to the low-power microprocessor, even superior to the near-threshold ARM Cortex-M3 design requiring $3.9 \times 10^3$ DMIPS/W \cite{dreslinski2013centip3de}.

%\begin{figure} [!t]
%    \centering
%    \includegraphics[scale=1.0]{../Figures/fig4.eps}
%    \caption{The FPGA-based verification platform for the prototype processor.}
%    \label{fig.fig1}
%\end{figure}

\begin{table}[!t]
\renewcommand{\arraystretch}{1.3}
\renewcommand{\tabcolsep}{2mm}
\centering
\caption{Implementation Results using CNTFET Ternary Gates}
\label{table:functions4}
\begin{tabular}{c|c|c|c}
\hline
\hline
Voltage & Total gates & Power & DMIPS/W\\ \hline
0.9V & 652 & 42.7 $\mu$W  & $3.06 \times 10^6$\\ \hline
\hline
\end{tabular}
\end{table}

\begin{table}[!t]
\renewcommand{\arraystretch}{1.3}
\renewcommand{\tabcolsep}{2mm}
\centering
\caption{Implementation Results using FPGA-based Ternary Logics}
\label{table:functions5}
\begin{tabular}{c|c|c|c|c|c}
\hline
\hline
Voltage & Frequency & ALMs & Registers & RAM & Power\\ \hline
0.9V & 150MHz & 803 & 339 & 9,216 bits & 1.09W \\ \hline
\hline
\end{tabular}
\end{table}

In order to validate the proposed ternary processor, we also implemented the ART-9 core in the FPGA-based verification platform.
For the practical implementation, all the ternary-based building blocks are emulated with the binary modules, adopting the binary-encoded ternary number system\cite{frieder1975algorithms}. 
Table V summarizes the implementation results of the binary-encoded ART-9 core, which utilizes only few hardware resources of Intel Stratix-V FPGA at the operating frequency of 150 MHz.
Targeting the dhrystone benchmark, the FPGA-based ART-9 core achieves 57.8 DMIPS/W by consuming 1.09W including two binary-encoded ternary memories.
As a result, the proposed frameworks successfully opens the preliminary results for realizing the ternary-based processor, which can be easily mapped to the future emerging ternary devices for allowing the extreme-low-power computing.

\section{Conclusion}
In this paper, we have proposed several designs and evaluation frameworks for developing ternary microprocessors, which are verified by the lightweight RISC-based ternary processor with 9-trit datapath.
Based on the balanced ternary number system, the proposed software-level framework efficiently supports a systematic way to construct assembly codes for the given ternary ISA.
Accepting the architecture-aware descriptions as well as the target technology information, the hardware-level framework is then followed to estimate several implementation-aware metrics, reducing the overall design overheads in the ternary number system. 
Based on the proposed frameworks, the fully-functional ART-9 microprocessor is developed and verified at different emerging technologies, offering attractive design methods for ternary processors.

%ART-9 ISA is effectively constructed with essential 24 instructions, each of which is encoded by a 9-trit length, by assuming one registerfile and two memory units like the typical RISC-based ISAs.
%The proposed 5-stage pipelined architecture successfully minimizes the number of stall cycles by adopting hazard-contro techniques.
%The verification tools are also introduced to construct the ternary-based assembly codes and to perform the logic-level simulations for the ART-9 core.
%We also provide implementation results of the ternary processor, for the first time, based on the reported CNTFET ternary gates as well as the FPGA-base verification platform, successfully offering the attractive solution for the next-generation microprocessor architecture.

\bibliographystyle{IEEEtran}
\bibliography{main.bbl}

% Generated by IEEEtran.bst, version: 1.14 (2015/08/26)
\begin{thebibliography}{10}
\providecommand{\url}[1]{#1}
\csname url@samestyle\endcsname
\providecommand{\newblock}{\relax}
\providecommand{\bibinfo}[2]{#2}
\providecommand{\BIBentrySTDinterwordspacing}{\spaceskip=0pt\relax}
\providecommand{\BIBentryALTinterwordstretchfactor}{4}
\providecommand{\BIBentryALTinterwordspacing}{\spaceskip=\fontdimen2\font plus
\BIBentryALTinterwordstretchfactor\fontdimen3\font minus
  \fontdimen4\font\relax}
\providecommand{\BIBforeignlanguage}[2]{{%
\expandafter\ifx\csname l@#1\endcsname\relax
\typeout{** WARNING: IEEEtran.bst: No hyphenation pattern has been}%
\typeout{** loaded for the language `#1'. Using the pattern for}%
\typeout{** the default language instead.}%
\else
\language=\csname l@#1\endcsname
\fi
#2}}
\providecommand{\BIBdecl}{\relax}
\BIBdecl

\bibitem{yeap2013smart}
G.~Yeap, ``Smart mobile $\text{SoC}$s driving the semiconductor industry:
  Technology trend, challenges and opportunities,'' in \emph{IEDM Tech. Dig.},
  Dec. 2013, pp. 1--3.

\bibitem{magen2004interconnect}
N.~Magen, A.~Kolodny, U.~Weiser, and N.~Shamir, ``Interconnect-power
  dissipation in a microprocessor,'' in \emph{Proc. Int. Workshop Syst, Level
  Interconnect Predict.}, 2004, pp. 7--13.

\bibitem{gaudet2016survey}
V.~Gaudet, ``A survey and tutorial on contemporary aspects of multiple-valued
  logic and its application to microelectronic circuits,'' \emph{IEEE J. Emerg.
  Sel. Topics Circuits Syst.}, vol.~6, no.~1, pp. 5--12, 2016.

\bibitem{lin2009novel}
S.~Lin, Y.-B. Kim, and F.~Lombardi, ``A novel {CNTFET}-based ternary logic gate
  design,'' in \emph{Proc. IEEE Int. Midwest Symp. Circuits Syst.}, Aug. 2009,
  pp. 435--438.

\bibitem{yang2012graphene}
H.~Yang \emph{et~al.}, ``Graphene barristor, a triode device with a
  gate-controlled schottky barrier,'' \emph{Science}, vol. 336, no. 6085, pp.
  1140--1143, 2012.

\bibitem{shin2015compact}
S.~Shin, E.~Jang, J.~W. Jeong, B.-G. Park, and K.~R. Kim, ``Compact design of
  low power standard ternary inverter based on {OFF}-state current mechanism
  using nano-{CMOS} technology,'' \emph{IEEE Trans. Electron Devices}, vol.~62,
  no.~8, pp. 2396--2403, 2015.

\bibitem{kim2018optimal}
S.~Kim, T.~Lim, and S.~Kang, ``An optimal gate design for the synthesis of
  ternary logic circuits,'' in \emph{Proc. 23rd Asia South Pacific Design
  Automat. Conf. (ASP-DAC)}, Jan. 2018, pp. 476--481.

\bibitem{kim2020logic}
S.~Kim, S.-Y. Lee, S.~Park, K.~R. Kim, and S.~Kang, ``{A logic synthesis
  methodology for low-power ternary logic circuits},'' \emph{IEEE Trans.
  Circuits Syst. I, Reg. Papers}, vol.~67, no.~9, pp. 3138--3151, 2020.

\bibitem{heo2018ternary}
S.~Heo \emph{et~al.}, ``Ternary full adder using multi-threshold voltage
  graphene barristors,'' \emph{IEEE Electron Device Lett.}, vol.~39, no.~12,
  pp. 1948--1951, Dec. 2018.

\bibitem{kang2017novel}
Y.~Kang \emph{et~al.}, ``A novel ternary multiplier based on ternary cmos
  compact model,'' in \emph{Proc. IEEE 47th Int. Symp. Multiple-Valued Log.
  (ISMVL)}, May. 2017, pp. 25--30.

\bibitem{choi2021design}
Y.~Choi, S.~Kim, K.~Lee, and S.~Kang, ``{Design and Analysis of a Low-Power
  Ternary SRAM},'' in \emph{Proc. IEEE Int. Symp. Circuits Syst. (ISCAS)}, Apr.
  2021, pp. 1--4.

\bibitem{PyconTernary}
\BIBentryALTinterwordspacing
``Ternary computer system,'' Accessed on: Sep. 17, 2021. [Online]. Available:
  \url{https://www.ternary-computing.com}
\BIBentrySTDinterwordspacing

\bibitem{DouglasW}
\BIBentryALTinterwordspacing
``The trillium architecture,'' Accessed on: Sep. 17, 2021. [Online]. Available:
  \url{https://homepage.divms.uiowa.edu/\texttildelow
  jones/ternary/trillium.shtml}
\BIBentrySTDinterwordspacing

\bibitem{narkhede2013design}
S.~Narkhede, G.~Kharate, and B.~Chaudhari, ``Design and implementation of an
  efficient instruction set for ternary processor,'' \emph{International
  Journal of Computer Applications}, vol.~83, no.~16, 2013.

\bibitem{RISC-V}
\BIBentryALTinterwordspacing
A.~Waterman, Y.~Lee, D.~A. Patterson, and K.~Asanovi{\'c}, ``{The RISC-V
  Instruction Set Manual, Volume I: User-Level ISA, Version 2.1},'' 2016.
  [Online]. Available: \url{https://riscv.org/specifications/}
\BIBentrySTDinterwordspacing

\bibitem{parhami2013arithmetic}
B.~Parhami and M.~McKeown, ``Arithmetic with binary-encoded balanced ternary
  numbers,'' in \emph{Proc. Asilomar Conf. Signals, Systems and Computers},
  2013, pp. 1130--1133.

\bibitem{8714897}
G.~Tagliavini, S.~Mach, D.~Rossi, A.~Marongiu, and L.~Benini, ``\text{Design}
  and \text{Evaluation} of \text{Small} \text{Float} \text{SIMD} extensions to
  the \text{RISC-V ISA},'' in \emph{Proc. of the Design, Automat. Test Eur.
  (DATE)}, 2019, pp. 654--657.

\bibitem{yiu2015definitive}
J.~Yiu, \emph{The Definitive Guide to ARM Cortex-M0 and Cortex-M0+
  Processors}.\hskip 1em plus 0.5em minus 0.4em\relax 2nd ed. Boca Raton, FL,
  USA: Academic, 2015.

\bibitem{schuiki2020stream}
F.~Schuiki \emph{et~al.}, ``{Stream semantic registers: A lightweight
  \text{RISC-V ISA} extension achieving full compute utilization in
  single-issue cores},'' \emph{IEEE Trans. Comput.}, vol.~70, no.~2, pp.
  212--227, 2020.

\bibitem{li2019reduce}
P.~Li, ``{Reduce Static Code Size and Improve RISC-V Compression},''
  \emph{Master’s thesis. EECS Department, Univ. of California, Berkeley},
  2019.

\bibitem{wang2012compression}
Y.~Wang \emph{et~al.}, ``A compression-based area-efficient recovery
  architecture for nonvolatile processors,'' in \emph{Proc. of the Design,
  Automat. Test Eur. (DATE)}, 2012, pp. 1519--1524.

\bibitem{zulehner2019matrix}
A.~Zulehner and R.~Wille, ``Matrix-vector vs. \text{M}atrix-matrix
  multiplication: Potential in \text{DD}-based simulation of quantum
  computations,'' in \emph{Proc. of the Design, Automat. Test Eur. (DATE)},
  2019, pp. 90--95.

\bibitem{york2002benchmarking}
R.~York, ``Benchmarking in context: Dhrystone,'' \emph{ARM, March}, 2002.

\bibitem{VexRiscv}
\BIBentryALTinterwordspacing
``Vexriscv,'' Accessed on: Sep. 17, 2021. [Online]. Available:
  \url{https://github.com/SpinalHDL/VexRiscv}
\BIBentrySTDinterwordspacing

\bibitem{PicoRV32}
\BIBentryALTinterwordspacing
``Picorv32,'' Accessed on: Sep. 17, 2021. [Online]. Available:
  \url{https://github.com/cliffordwolf/picorv32}
\BIBentrySTDinterwordspacing

\bibitem{dreslinski2013centip3de}
R.~G. Dreslinski \emph{et~al.}, ``{Centip3de: A 64-core, 3d stacked
  near-threshold system},'' \emph{IEEE Micro}, vol.~33, no.~2, pp. 8--16, 2013.

\bibitem{frieder1975algorithms}
G.~Frieder and C.~Luk, ``Algorithms for binary coded balanced and ordinary
  ternary operations,'' \emph{IEEE Trans. Comput.}, vol. 100, no.~2, pp.
  212--215, 1975.

\end{thebibliography}

\end{document}